\documentstyle{article}
\title{
ACCRETION DISC THEORY: FROM THE STANDARD MODEL UNTIL ADVECTION}

\author{G.S.BISNOVATYI-KOGAN\thanks{
           IKI, Profsoyuznaya 84/32, Moscow 117810
           Russia, E-mail: gkogan@mx.iki.rssi.ru}}
\date{}
\begin{document}
\maketitle
\begin{abstract}
Accretion disc theory was first developed as a theory with
the local heat balance, where the whole energy produced by a viscous
heating was emitted to the sides of the disc. One of the most important
new invention of this theory was a phenomenological treatment of the
turbulent viscosity, known as ''alpha'' prescription, when the (r$\phi$)
component of the stress tensor was approximated by ($\alpha$ P) with a unknown
constant $\alpha$. This prescription played the role in the accretion
disc theory as well important as the mixing-length theory of convection
for stellar evolution. Sources of turbulence in the accretion disc are
discussed, including nonlinear hydrodynamical turbulence, convection
and magnetic field role. In parallel to the optically thick
geometrically thin accretion disc models, a new branch of the optically
thin accretion disc models was discovered, with a larger thickness
for the same total luminosity. The choice between these solutions
should be done of the base of a stability analysis.
The ideas underlying the necessity to include advection into the
accretion disc theory are presented and first models with advection
are reviewed. The present status of the solution for a low-luminous
optically thin accretion disc model with advection is discussed and
the limits for an advection dominated accretion flows
(ADAF) imposed by the presence of magnetic field are analysed.
\end{abstract}

\section{Introduction}
Accretion is served as a source of energy in many astrophysical
objects, including different types of binary stars, binary X-ray sources,
most probably quasars and active galactic nuclei (AGN).
While first developement
of accretion theory started long time ago (Bondi and Hoyle, 1944;
Bondi, 1952), the intensive
developement of this theory began after discovery of first X ray sources
(Giacconi et al, 1962) and quasars (Schmidt, 1963). Accretion into stars,
including neutron stars, is ended by a collision with an inner boundary,
which may be a stellar surface, or outer boundary of a magnetosphere
for strongly magnetized stars. We may be sure in this case, that all
gravitational energy of the falling matter will be transformed into heat
and radiated outward.

Situation is quite different for sources containing black holes, which are
discovered in some binary X-ray sources in the galaxy, as well as in many
AGN. Here matter is falling to the horizon, from where no radiation arrives,
so all luminocity is formed on the way to it. The efficiency of accretion
is not known from the beginning, contrary to the accretion into a star,
and depends strongly on such factors, like angular momentum of the
falling matter, and magnetic field embedded into it. It was first
shown by Schwarzman (1971), that during spherical accretion of
nonmagnetized gas the efficiency may be as small as $10^{-8}$ for
sufficiently low mass fluxes. He had shown that presence of magnetic field in
the accretion flux matter increase the efficiency up to about $10\%$, and
account of heating of matter due to magnetic field annihilation in
the flux rises the efficiency up to about $30\%$ (Bisnovatyi-Kogan,
Ruzmaikin, 1974). In the case of a thin disc accretion when matter has
large angular momentum, the efficiency is about $1/2$ of the efficiency
of accretion
into a star with a radius equal to the radius of the last stable orbit.
Matter cannot emit all the gravitational energy, part of which is
absorbed by the black hole. In the case of geometrically thick and
optically thin accretion discs the situation is approaching the case
of spherical symmetry, and a presence of a magnetic field playes also
a critical role.

Here we consider a developement of the theory of a disk accretion,
starting from creation of a so called "standard model",
and discuss recent trends, connected with a presence of advection.

\section{Developement of the standard model of the disc accretion into
        a black hole}

Matter falling into a black hole is gathered into a disc when its
angular momentum is sufficiently high. It happens when the matter falling
into a black hole comes from the neighbouring
ordinary star companion in the binary, or when the matter appears as a
result of a tidal disruption of the star which trajectory of motion
approaches sufficiently close to the black hole, so that forces of
selfgravity could be overcomed. The first situation is observed in
many galactic X-ray sources containing a stellar mass black hole
(Cherepashchuk, 1996). A tidal disruption happens in quasars and
active galactic nuclei (AGN), if the model of supermassive black
hole surrounded by a dense stellar cluster
of Lynden-Bell (1969) is true for these objects.

The models of the accretion disc structure around a black hole had
been investigated by Lynden-Bell (1969), Pringle and Rees (1972).
The modern "standard" theory of the disc accretion was formulated in
the papers of Shakura (1972), Novikov and Thorne (1973) and
Shakura and Sunyaev (1973). It is important to note, that all authors
of the accretion disc theory from USSR were students (N.I.Shakura)
or collaborators (I.D.Novikov and R.A.Sunyaev) of academician
Ya.B.Zeldovich, who was not among the authors, but whose influence
on them hardly could be overestimated.

The equations of the standard disc accretion theory were first formulated
by Shakura (1972); some corrections and generalization to general
relativity (GR) were done by Novikov and Thorne (1973), see also correction
to their equations in GR made by Riffert \& Herold (1995).
The main idea of this theory is to describe a
geometrically thin non-self-gravitating disc of the mass $M_d$, which is much
smaller then the mass of the black hole $M$, by hydrodynamic
equations averaged over the disc thickness $2h$.

\subsection{Equilibrium equations}

The small thickness of the disc in comparison with its radius $h \ll r$
indicate to small importance of the pressure gradient
$\nabla P$ in comparison with
gravity and inertia forces. That leads to a simple
radial equilibrium equation  denoting the balance between the last two
forces occuring when the angular velocity of the disc $\Omega$ is equal to
the Keplerian one $\Omega_K$,

\begin{equation}
\label{ref1.1}
\Omega=\Omega_K=\left(\frac{GM}{r^3}\right)^{1/2}.
\end{equation}
Note, just before a last stable orbit around a black hole, and of course
inside it, this suggestion fails, but in the "standard"
accretion disc model the relation (\ref{ref1.1}) is suggested to be fulfilled
all over the disc, with an inner boundary at the last stable orbit.

The equilibrium equation in the vertical $z$-direction is determined by a
balance between the gravitational force and pressure gradient

\begin{equation}
\label{ref1.2}
\frac{dP}{dz}=-\rho\frac{GMz}{r^3}
\end{equation}
For a thin disc this differential equation is substituted by an
algebraic one, determining the half-thickness of the disc in the form

\begin{equation}
\label{ref1.3}
h \approx \frac{1}{\Omega_K}\left(2\frac{P}{\rho}\right)^{1/2}.
\end{equation}
The balance of angular momentum, related to the $\phi$ component of the
Euler equation has an integral in a stationary case written as

\begin{equation}
\label{ref1.4}
\dot M(j-j_{in})=-2\pi r^2\,2ht_{r\phi},\quad t_{r\phi}=
\eta r\frac{d\Omega}{dr}.
\end{equation}
Here $j=v_{\phi}r=\Omega r^2$ is a specific angular momentum,
$t_{r\phi}$ is a component of the viscous stress tensor, $\dot M>0$ is a mass
flux per unit time into a black hole,
$j_0$ is an integration constant having,
after multiplication by $\dot M$, a physical
sence of difference between viscous and advective flux of the
angular momentum, when $j_{in}$ itself is equal to the specific angular
momentum of matter
falling into a black hole. In the standard theory the value of $j_{in}$
is determined separately, from physical considerations.
For the accretion
into a black hole it is suggested, that on the last stable orbit the
gradient of the angular velocity is zero, corresponding to zero
viscous momentum flux. In that case

\begin{equation}
\label{ref1.5}
j_{in}=\Omega_K r_{in}^2,
\end{equation}
corresponding to the Keplerian angular momentum of the matter on the last
stable orbit. During accretion into a slowly rotating star which
angular velocity is smaller than a Keplerian velocity on the inner edge
of the disc, there is a maximum of the angular velocity close to its surface,
where viscous flux is zero, and there is a boundary layer between
this point and stellar surface. In that case (\ref{ref1.5}) remains to
be valid. The situation is different for accretion discs around rapidly
rotating stars with a critical Keplerian speed on the equator. Here
there is no extremum of the agular velocity of the disc which smoothly
joins the star. In stationary self-consistent situation when the accreting
star remains to rotate critically during the process of a disc accretion, the
specific angular momentum of matter joining the star is determined by a
relation (Bisnovatyi-Kogan, 1993):
$j_{in}=\frac{dJ}{dM}\vert_{crit}$,
where the derivative is taken along the states of the star having a Keplerian
equatorial speed. For stars with a polytropic structure, corresponding to
equation of state $P=K\rho^{1+\frac{1}{n}}$, this derivative
is calculated numerically
giving the value $j_{in}=0.176 \Omega_K r_{in}^2$ for $n=1.5$;
$0$ for $n=2.5$; and negative values of $j_{in}$ for larger $n$.

Note, that in the pioneering paper of Shakura (1972) the integration
constant $j_{in}$ was found as in (\ref{ref1.5}), but was taken
zero in his subsequent formulae. Importance of
using $j_{in}$ in the form (\ref{ref1.5}) was noticed by Novikov and
Thorne (1973), and became a feature of the standard model.

\subsection{Viscosity}

The choice of the viscosity  coefficient is the most difficult and
speculative problem of the accterion disc theory. In the laminar case of
microscopic (atomic or plasma) viscosity, which is very low, the stationary
accretion disc must be very massive and very thick, and before its formation
the matter is collected by disc leading to a small flux inside.
It contradicts to observations of X-ray binaries,
where a considerable matter flux along the accretion disc may be explained
only when viscosity coefficient is much larger then the microscopic one.
In the paper of Shakura (1972) it was suggested, that matter in the disc
is turbulent, what determines a turbulent viscous stress tensor,
parametrized by a pressure

\begin{equation}
\label{ref1.7}
t_{r\phi}=-\alpha\rho v_s^2 = -\alpha P,
\end{equation}
where $v_s$ is a sound speed in the matter.
This simple presentation comes out from a relation for a turbulent viscosity
coefficient $\eta_t\approx \rho v_t l$ with an average turbulent velocity
$v_t$ and mean free path of the turbulent element $l$. It follows from
the definition of $t_{r\phi}$ in (\ref{ref1.4}), when we take $l \approx h$
from (\ref{ref1.3})

\begin{equation}
\label{ref1.8}
t_{r\phi}=\rho v_t h r \frac{d\Omega}{dr} \approx \rho v_t v_s =-\alpha
\rho v_s^2,
\end{equation}
where a coefficient $\alpha<1$ is connecting the turbulent and sound speeds
$v_t=\alpha v_s$. Presentations of $t_{r\phi}$ in (\ref{ref1.7}) and
(\ref{ref1.8}) are equivalent, and only when the angular velocity
differs considerably from the Keplerian one the first relation to the
right in (\ref{ref1.8}) is more preferable. That does not appear
(by definition) in the standard theory, but may happen when advective
terms are included.

Developement of a turbulence in the acctretion disc cannot be justified
simply, because a Keplerian disc is stable in linear approximation to
the developement of perturbations. It was suggested by Ya.B.Zeldovich,
that in presence of very large Reynolds number
${\rm Re}=\frac{\rho v l}{\eta}$
the amplitude of perturbations at which nonlinear effects become important
is very low, so in this situation a turbulence may develope due to
nonlinear instability even when the disc is stable in linear approximation.
Another source of viscous stresses may arise from a magnetic field,
but it was suggested by Shakura (1972), that magnetic stresses cannot
exceed the turbulent ones.

Magnetic plasma instability as a source of the turbulence in the
accretion discs has been studied extensively in last years (see
review of Balbus and Hawley, 1998). They used an instability of the
uniform magnetic field parallel to the axis in differentially rotating
disc, discovered by Velikhov (1959). It could be really important in
absence of any other source of the turbulence, but it is hard to belive
that there is no radial or azimuthal component of the magnetic field
in matter flowing into the accretion disc from the companion star.
In that case the field amplification due to twisting by a differential
rotation take place without necessity of any kind of instability.

It was shown by Bisnovatyi-Kogan and Blinnikov (1976, 1977), that inner
regions of a highly luminous accretion discs where pressure is dominated
by radiation, are unstable to vertical convection. Developement of this
convection produce a turbulence, needed for a high viscosity. Other
regions of a standard accretion disc should be stable to developement of a
vertical convection, so other ways of a turbulence exitation are needed
there. With alpha- prescription of viscosity the equation of angular
momentum conservation is written in the plane of the disc as

\begin{equation}
\label{ref1.9}
\dot M(j-j_{in})=
4\pi r^2 \alpha P_0 h.
\end{equation}
When angular velocity is far from Keplerian the relation
(\ref{ref1.4}) is valid with a coefficient of a turbulent viscosity

\begin{equation}
\label{ref1.10}
\eta=\alpha\rho_0 v_{s0} h,
\end{equation}
where values with the index "0" denote the plane of the disc.

\subsection{Heat balance}

In the standard theory a heat balance is local, what means that all
heat produced by viscosity in the ring between $r$ and $r+dr$ is
radiated through the sides of disc at the same $r$. The heat production
rate $Q_+$ related to the surface unit of the disc is written as

\begin{equation}
\label{ref1.11}
Q_+=h\,t_{r\phi}r\frac{d\Omega}{dr}=\frac{3}{8\pi}\dot M \frac{GM}{r^3}
\left(1-\frac{j_{in}}{j}\right).
\end{equation}
Heat losses by a disc depend on its optical depth. The first standard disc
model of Shakura (1972) considered a geometrically thin disc as an optically
thick in a vertical direction. That implies enegry losses $Q_-$ from the disc
due to a radiative conductivity, after a substitution of
the differential eqiation
of a heat transfer by an algebraic relation

\begin{equation}
\label{ref1.12}
Q_- \approx \frac{4}{3} \frac{acT^4}{\kappa \Sigma}.
\end{equation}
Here $a$ is a constant of a radiation energy density,
$c$ is a speed of light,
$T$ is a temperature in the disc plane, $\kappa$ is a matter opacity,
and a surface density
$\Sigma=2\rho h$.
Here and below
$\rho,\, T,\,P$ without the index "0" are related to the disc plane.
The heat balance equation is represented by a relation

\begin{equation}
\label{ref1.14}
Q_+=Q_-,
\end{equation}
A continuity equation in the standard model of the stationary accretion flow
is used for finding of a radial velocity $v_r$

\begin{equation}
\label{ref1.14a}
v_r=\frac{\dot M}{4\pi rh\rho}=\frac{\dot M}{2\pi r\Sigma}.
\end{equation}
Equations (\ref{ref1.1}),(\ref{ref1.3}),(\ref{ref1.9}),
(\ref{ref1.14}), completed by an equation of state $P(\rho,T)$ and relation
for the opacity
$\kappa=\kappa(\rho, T)$ represent a full set of equatiions
for a standard disc model. For power low equations of state of an ideal gas
$P=P_g=\rho {\cal R} T$ (${\cal R}$ is a gas constant), or radiation pressure
$P=P_r=\frac{aT^4}{3}$, and opacity in the form of electron scattering
$\kappa_e$, or Karammers formulae $\kappa_k$,
the solution of a standard disc accretion theory is obtained
analytically (Shakura, 1972; Novikov, Thorne, 1973; Shakura, Sunyaev, 1973).
Checking the suggestion of a large optical thickness confirms a
self-consistency of the model. One of the shortcoming of the analytical
solutions of the standard model lay in the fact, that
solutions for different regions
of the disc with different equation of states and opacities are not matched
to each other.

\subsection{Optically thin solution}

Few years after appearence of the standard model it was found that
in addition to the opically thick disc solution there is another branch
of the solution for the disc structure with the same input parameters
$M,\,\dot M,\,\alpha$ which is also self-consistent and has a small
optical thickness (Shapiro, Lightman, Eardley, 1976). Suggestion of
the small optical thickness implies another equation of energy losses,
determined by a volume emission
$Q_- \approx q\, \rho\,h$,
where due to the Kirghoff law the emissivity of the unit of a volume $q$ is
connected with a Plankian averaged opacity
$\kappa_p$ by an approximate relation
$q \approx acT_0^4 \kappa_p$.
Note, that Krammers formulae for opacity
are obtained after Rosseland averaging of the
frequency dependent absorption coefficient.
In the optically thin limit the
pressure is determined by a gas $P=P_g$. Analytical solutions are obtained
here as well, from the same equations with volume losses and gas pressure.
In the optically thin solution the thickness of the disc is larger then
in the optically thick one, and density is lower.

While heating
by viscosity is determined mainly by heavy ions, and cooling is determined
by electrons, the rate of the energy exchange between them is important for
a thermal structure of the disc. The energy balance equations are written
separately for ions and electrons. For small accretion rates and lower
matter density the rate of energy exchange due to binary collisions is
so slow, that in the thermal balance the ions are much hotter then the
electrons. That also implies a high disc thickness and brings the
standard accretion theory to the border of its applicability.
Nevertheless, in the highly turbulent plasma the energy exchange
between ions and electrons may be strongly enhanced due to presence
of fluctuating electrical fields, where electrons and ions gain the
same energy. In such conditions difference of temperatures between ions and
electrons may be negligible. Regretfully, the theory of relaxation
in the turbulent plasma is not completed, but there are indications
to a large exhancement of the relaxation in presence of plasma turbulence,
in comparison with the binary collisions (Quataert, 1997).

\subsection{
Accretion disc structure from equations describing continuously
optically thin and optically thick disc regions}

In order to find equations of the disc structure
 valid in both limiting cases of optically thick and optically thin disc,
and smootly describing transition between them, Eddington
approximation had been used
for obtaining formulae for a heat flux and for a radiation
pressure (Artemoma et al., 1996).
The following expressions had been obtained
for the vertical energy flux from the disc
$F_0$, and the radiation pressure in the
symmetry plane

\begin{equation}
\label{ref11.15}
 F_{0}={2acT_0^4 \over 3\tau_{0}}\left(1+{4 \over 3\tau_{0}}+
 {2 \over 3\tau_{*}^2}\right)^{-1},\quad
 P_{rad,0}={aT_0^4 \over 3}
 {1+{4 \over 3 \tau_{0}}\over 1+{4 \over 3\tau_{0}}+
 {2\over 3\tau_{*}^2}},
\end{equation}
where
 $ \tau_{0}=\kappa_e \rho h$,
  $\tau_{*}=\left(\tau_{0}\tau_{\alpha 0}\right)^{1/2}$,
  $\tau_{\alpha 0} \approx \kappa_p \rho h$.
At $\tau_0 \gg \tau_* \gg 1$ we have (\ref{ref1.12}) from (\ref{ref11.15}).
In the optically thin limit $\tau_* \ll \tau_0 \ll 1$ we get

\begin{equation}
\label{ref11.17}
 F_{0}=acT_0^4 \tau_{\alpha 0}, \quad
 P_{rad,0}={2 \over 3}acT_0^4 \tau_{\alpha 0}.
\end{equation}
Using $F_0$ instead of $Q_-$ and equation of state
$P=\rho {\cal R} T+P_{rad,0}$,
the equations of accretion disc structure together with equation
$Q_+=F_0$,
with $Q_+$ from (\ref{ref1.11}),
have been solved numerically by Artemova et al. (1996).
It occures that two solutions, optically thick and optically thin, exist
separately when luminosity is not very large. Two solutions intersect at
$\dot m=\dot m_b$ and there is no global solution for accretion disc at
$\dot m > \dot m_b$ (see Fig.1). It was concluded by Artemova et al (1996),
that in order to obtain a global physically meaningful solution
at $\dot m > \dot m_b$, account of advection is needed.

\begin{figure}
\vspace{8cm}  
\caption{
 The dependences of the optical depth $\tau_0$ on
radius, $r_*=r/r_{g}$, for the case $M_{BH}=10^8\;M_\odot$, $\alpha=1.0$
and different values of $\dot m$. The thin solid, dot-triple dash,
long dashed, heavy solid, short dashed, dotted and dot-dashed curves
correspond to $\dot m=1.0, 3.0, 8.0, 9.35, 10.0, 11.0, 15.0$,
respectively. The upper curves correspond to the optically thick
family, lower curves correspond to the optically thin
family.}
\end{figure}

\section{Accretion discs with advection}

Standard model gives somewhat nonphysical behaviour near the inner
edge of the accretion disc around a black hole.
For high mass fluxes when central regions are radiation-dominated
($P \approx P_r, \,\, \kappa \approx \kappa_e$), the radial dependence
follows relations (Shakura, Sunyaev, 1973)

\begin{equation}
\label{ref31.1}
\rho \sim r^{3/2}{\cal J}^{-2} \rightarrow \infty,
\quad T \sim r^{-3/8},
\end{equation}
$$\quad h \sim {\cal J} \rightarrow 0,
\quad \Sigma \sim r^{3/2}{\cal J}^{-1} \rightarrow \infty,
\quad v_r \sim r^{-5/2}{\cal J} \rightarrow 0,
$$
where limits relate to the inner edge of the disc with $r=r_{in}$,
${\cal J}=1-\frac{j_{in}}{j}=1-\sqrt{\frac{r_{in}}{r}}$.
At smaller $\dot M$, when near the inner edge
$P \approx P_g, \,\, \kappa \approx \kappa_e$, there are different type
of singularities

\begin{equation}
\label{ref31.3}
\rho \sim r^{-33/20}{\cal J}^{2/5} \rightarrow 0,
\quad T \sim r^{-9/10}{\cal J}^{2/5}\rightarrow 0,
\end{equation}
$$\quad h \sim r^{21/20}{\cal J}^{1/5} \rightarrow 0,
\quad \Sigma \sim r^{-3/5}{\cal J}^{3/5} \rightarrow 0,
\quad v_r \sim r^{-2/5}{\cal J}^{-3/5} \rightarrow \infty.
$$
This results from the local
form of the equation of the thermal balance (\ref{ref1.14}). It is
clear from physical ground, that when a local heat production due
to viscosity goes to zero, the heat brought by radial motion of matter
along the accretion disc becomes more important. In presence of
this advective heating (or cooling term,
depending on the radial entropy $S$ gradient) written as

\begin{equation}
\label{ref3.1}
Q_{adv}=\frac{\dot M}{2\pi r}T \frac{dS}{dr},
\end{equation}
the equation of a heat balance is modified to
$Q_+ + Q_{adv}=Q_-$.
In order to describe self-consistently the structure of the accertion disc
we should also modify the radial disc equilibrium, including pressure
and inertia terms

\begin{equation}
\label{ref3.3}
r(\Omega^2-\Omega_K^2)=\frac{1}{\rho}\frac{dP}{dr}-v_r\frac{dv_r}{dr}.
\end{equation}
Appearence of inertia term leads to transonic radial flow with a
singular point. Conditions of a continious passing of the solution through
a critical point choose a unique value of the integration constant $j_{in}$.
First approximate solution for the advective disc structure have been
obtained by Paczynski and Bisnovatyi-Kogan (1981), but a corresponding set
of equations had beed discussed earlier (Hoshi and Shibazaki, 1977;
Liang and Thompson, 1980). Attempts to find a solution for advective
disc structure (see e.g. Matsumoto et al., 1984; Abramovicz et al., 1988)
gave the following results. For moderate values of $\dot M$ a unique
continuous transonic solution was found, passing through singular points,
and corresponding to a unique value of $j_{in}$.
The number of critical ponts in the radial flow happens always to be
more then unity. This is connected with two reasons. First, the
gravitational potential $\phi_g$ in papers dealing with advective disc
solutions was different from newtonian one (Paczi\'nski and Wiita, 1980):
$\phi_g = \frac{GM}{r-r_g}, \quad r_g=\frac{2GM}{c^2}$.
The advantage of this potential is a realistic approximation of the
general relativistic (GR) effects, namely, infinitive gravitational attraction
at a gravitational radius $r_g$, and existence of the stable circular
orbits only up to $r=3r_g$, like in exact GR. Appearence of two critical
points for a radial flow in this potential was analysed by Chakrabarti and
Molteni (1993). The second reason of multiplicity of singular points
is connected with using of equations averaged over a thickness of the disc.
That changes a structure of hydrodynamic equations, leading to a position
of singular points not coinsiding with a unit Mach number point, and
increasing a number of critical points.

When $\dot M$ is becoming so high, that radiation pressure starts to be
important, still unresolved problems appear in a construction of the
advective disc model. These problems are connected with increasing of
a number of a critical points from one side, and loss of uniqueness
of the transonic solution from another. So, with increasing of $\dot M$
the solution becomes nonunique at some parameters,
or was not found at all
(see Matsumoto et al., 1984; Abramovicz et al., 1988;
Artemova et al., 1996a). At high $\dot M$ the integral curves are very
sensitive to input conditions: form of viscosity stresses (\ref{ref1.4})
or (\ref{ref1.7}), choice of boundary conditions etc. The system of
equations has a very small resource of stability, so it cannot be
excluded, that the failures are connected with an
inproper choice of a numerical
method and developement of numerical instabilities prevents of finding
a unique physical solution. In addition to continuous solutions,
solutions with standing shock waves have been investigated
(Chakrabarti, 1996).

\subsection{Two-temperature advective discs}

In the optically thin accretion discs at low mass fluxes the density
of the matter is low and energy exchange between electrons and ions
due to binary collisions is slow. In this situation, due to different
mechanisms of heating and cooling for electrons and ions, they may
have different temperatures. First it was realized by Shapiro,
Lightman, Eardley (1976) where advection was not included. It was
noticed by Narayan and Yu (1995), that advection in this case is becoming
extremely important. It may carry the main energy flux into a black
hole, leaving rather low efficiency of the accretion up to $10^{-3}\,-\,
10^{-4}$ (advective dominated accretion flows - ADAF).
This conclusion is valid only when the effects, connected
with magnetic field annihilation and heating of matter due to it are
neglected.

  In the ADAF solution the ion temperature is about a virial one
$kT_i \sim GMm_i/r$, what means that even at high initial angular
momentum the disc becomes very thick, forming practically a quasi-spherical
accretion flow. It is connected also with an "alpha" prescription of
viscosity. At high ion temperatures, connected with a strong
viscous heating, the ionic pressure becomes high, making
the viscosity very effective.
So, due to suggestion of "alpha" viscosity in the situation,
when energy losses by ions are very low, some kind of a "thermo-viscous"
instability is developed, because heating increases a viscosity, and
viscosity increases a heating. Developement of this instability
leads to formation of ADAF.

A full account of the processes, connected with a presence of magnetic
field in the flow, is changing considerably the picture of ADAF. It was
shown by Schwarzman (1971), that radial component of the magnetic field
increses so rapidly in the spherical flow, that equipartition between
magnetic and kinetic energy is reached in the flow far from the black hole
horizon. In the region where the main enegry prodiction takes place, the
condition of equipartition takes place. In presence of a high magnetic
field the efficiecy of a radiation during accretion of an interstellar matter
into a black hole increase enormously from $\sim 10^{-8}$ up to $\sim 0.1$
(Schwarzman, 1971), due to efficiency of a magneto-bremstrahlung
radiation. So possibility of ADAF regime for  a spherical accretion was
noticed long time ago. To support the condition of equipartition a
continuous magnetic field reconnection is necessary, leading to annihilation
of the magnetic flux and heating of matter due to Ohmic heating. It was
obtained by Bisnovatyi-Kogan and Ruzmaikin (1974), that due to Ohmic heating
the efficiency of a radial
accretion into a black hole may become as high as
$\sim 30\%$. The rate of the Ohmic heating in the condition of
equipartition was obtained in the form

\begin{equation}
\label{ref3.5}
T\frac{dS}{dr} = -\frac{3}{2}\frac{B^2}{8\pi \rho r}.
\end{equation}
In the supersonic flow of the radial acrretion equipartition between
magnetic and kinetic energy was suggested by Schwarzman (1971):

\begin{equation}
\label{ref3.6}
\frac{B^2}{8\pi} \approx \frac{\rho v_r^2}{2}=\frac{\rho GM}{r}.
\end{equation}
For the disc accretion, where there is more time for a field dissipation,
almost equipartition was suggested (Shakura, 1972) between magnetic and
turbulent energy, what reduces with account of "alpha" prescription
of viscosity to a relation

\begin{equation}
\label{ref3.7}
\frac{B^2}{8\pi}  \sim \frac{\rho v_t^2}{2} = \frac{3}{2}\alpha_m^2 P,
\end{equation}
where $\alpha_m$ characterises a magnetic viscosity in a way similar to
the turbulent $\alpha$ viscosity. It was suggested by Bisnovatyi-Kogan and
Ruzmaikin (1976) the similarity between viscous and magnetic Reynolds
numbers, or between turbulent and magnetic viscosity coefficients

\begin{equation}
\label{ref3.8}
Re=\frac{\rho v l}{\eta}, \quad Re_m=\frac{\rho v l}{\eta_m},
\end{equation}
where the turbulent magnetic viscosity $\eta_m$ is connected with a turbulent
conductivity
$\sigma=\frac{\rho c^2}{4\pi \eta_m}$.
Taking $\eta_m=\frac{\alpha_m}{\alpha}\eta$, we get a turbulent conductivity

\begin{equation}
\label{ref3.10}
\sigma=\frac{ c^2}{4\pi \alpha_m h v_s}, \quad v_s^2=\frac{P_g}{\rho}
\end{equation}
in the optically thin discs. For the radial accretion the turbulent
conductivity may contain mean free path of a turbulent element $l_t$,
and turbulent viscocity $v_t$ in (\ref{ref3.10}) instead of
$h$ and $v_s$.
In ADAF solutions, where ionic temperature is of the order of the virial one
the two suggestions (\ref{ref3.6}) and (\ref{ref3.7}) almost coinside at
$\alpha_m \sim 1$.

The heating of the matter due to an Ohmic dissipation
may be obtained from the Ohm's law for a radial accretion in the form

\begin{equation}
\label{ref3.11}
T\frac{dS}{dr} =-\frac{ \sigma {\cal E}^2}{v_r}
\approx -\sigma \frac{v_t^2 B^2}{v_r c^2}
=-\frac{B^2 v_t}{4 \pi \alpha_m v_r l_t},
\end{equation}
what coinsides with (\ref{ref3.5})
when $ \alpha_m=\frac{4 r v_t}{3 v_r l_t}$, or $l_t
=\frac{4 r v_t}{3 v_r \alpha_m}$.
Here a local electrical field strength
in a highly conducting plasma is of the
order of ${\cal E} \sim \frac{v_t B}{c}$ for the radial accretion.

Equations for a radial temperature dependence in the accretion disc,
separate for the ions and electrons are written as (Bisnovatyi-Kogan
and Lovelace, 1997)

\begin{equation}
\label{ref3.12a}
{dE_i\over dt}-{P_i\over\rho^2}{d\rho\over dt}=
{\cal H}_{\eta i}+{\cal H}_{Bi}-Q_{ie}~,
\end{equation}
\begin{equation}
\label{ref3.12b}
{dE_e\over dt}-{P_e\over\rho^2}{d\rho\over dt}=
{\cal H}_{\eta e}+{\cal H}_{Be}+Q_{ie}-{\cal C}_{brem}-{\cal C}_{cyc}~,
\end{equation}
Here $\frac{d}{dt}=\frac{\partial}{\partial t}
+v_r\frac{\partial}{\partial r}$.
A rate of a viscous heating of ions ${\cal H}_{\eta i}$ is obtained from
(\ref{ref1.11}) as

\begin{equation}
\label{ref3.13}
{\cal H}_{\eta i}=\frac{2\pi r}{\dot M}Q_+ \,=\,\frac{3}{2}
\alpha\frac{v_K v_s^2}{r}, \quad {\cal H}_{\eta e}=
\sqrt{\frac{m_e}{m_i}} {\cal H}_{\eta i},
\end{equation}
where $v_K=r\Omega_K$. The rate of the energy exchange between ions and
electrons due to the binary collisions was obtained by Landau (1937),
Spitzer (1940) as

\begin{equation}
\label{ref3.15}
Q_{ie} \approx
{4(2\pi)^{1\over2}n e^4 \over m_im_e}
\left({T_e\over m_e}+{T_i\over m_i}\right)^{-{3\over2}}
{\ell}n\Lambda(T_i-T_e),
\end{equation}
with $\ell n \Lambda={\cal O}(20)$ the Coulomb logarithm.
The electron bremstrahlung ${\cal C}_{brem}$ and magneto-bremstrahlung
${\cal C}_{cyc}$ cooling are taken into account.
The expression for an Ohmic heating in the turbulent accretion disc
may be written in different ways, using different velocities $v_E$ in the
expression for an effective electrical field
${\cal E}=\frac{v_E B}{c}$.
A self-consistency of the model requires, that expressions
for a magnetic heating of the matter ${\cal H}_B$, obtained from the
condition of stationarity of the flow (\ref{ref3.5}), and from the Ohm's
law (\ref{ref3.11}), should be identical. That gives some restrictions
for the choice of a characteristic velocity $v_E$. Comparison
between (\ref{ref3.5}) and (\ref{ref3.11}) shows the identity of these two
expressions at
$v_E=v_r, \quad, \frac{\alpha}{{\cal J}\alpha_m}=\frac{3\sqrt{2}}{4}$.
So, the model is becoming self-consistent at the reasonable choice of
the parameters. Note, that in the advective models
${\cal J}$ is substituted by another function which is
not zero at the inner edge of the disc. The heating due to magnetic
field reconnection ${\cal H}_B$ in the equations (\ref{ref3.12a}),
(\ref{ref3.12b}), may be written as

\begin{equation}
\label{ref3.23}
{\cal H}_B=\frac{3}{16 \pi}\frac{B^2}{r \rho} v_r
    = \frac{1}{2{\cal J}}{\cal H}_{\eta i}\left(\frac{v_B}{v_K}\right).
\end{equation}
So, at $v_B=v_K$ the expressions for viscous and magnetic heating are
almost identical.
The distribution of the magnetic heating between electrons and ions has
a critical influence on the model, if we neglect the influence of a plasma
turbulence on the energy relaxation, and take into account only
the energy exchange by binary collisions from (\ref{ref3.15}). Observations
of the magnetic field reconnection in the solar flares show (Tsuneta, 1996),
that electronic heating prevails.

It follows from the physical picture of
the field reconnection, that transformation of the magnetic energy into
a heat is connected with the change of the magnetic flux, generation
of the vortex electrical field, accelerating the particles. This vortex
field has a scale of the turbulent element and suffers rapid and
chaotic changes. The accellerating forces on electrons and protons in
this fields are identical, but accelerations themselfs differ $\sim 2000$
times, so during a sufficiently short time of the tubulent pulsation
the electron may gain much larger energy, then the protons.
Additional particle
acceleration and heating  happens on the shock fronts, appearing around
turbulent cells, where reconnection happens. In this process
acceleration of the electrons is also more effective than of the protons.
In the paper of Bisnovatyi-Kogan and Lovelace (1997) the equations
(\ref{ref3.12a}), (\ref{ref3.12b})
have been solved in the approximation of nonrelativistic
electrons, $v_B$=$v_K$, what permitted to unite a viscous and magnetic
heating into a unique formila. The combined heating of the electrons
and ions were taken as
${\cal H}_e=(2-g){\cal H}_{\eta i},\quad {\cal H}_e =g{\cal H}_{\eta i}$.
In the expression for a cyclotron emission self-absorption was taken into
account according to Trubnikov (1973). The results of calculations for
$g=0.5 \div 1$ show that almost all energy of the electrons is radiated,
so the relative efficiency of the two-temperature, optically thin disc
accretion cannot become lower then 0.25. Note again that accurate account
of a plasma turbulence for a thermal relaxation and corresponding increase
of the term $Q_{ie}$ may restore the relative efficiency to its unity value,
corresponding to the optically thick discs.

\section{Discussion}

Observational evidences for existence of black holes inside our Galaxy
and in the active galactic nuclei (Cherepashchuk, 1996, Ho, 1998)
make necessary to revise theoretical models of the disc accretion.
Large part of high energy radiation indicates to its origin
close to the black hole, where standard accretion disc model is not a
appropriate. The improvements of a model are connected with account of
advective terms and more accurate treatement of the magnetic field effects.
Conclusions about existence of ADAF solution for an optically thin
accretion disc at low mass flux are connected with an incomplete account
of the effects connected with magnetic field annihilation. Their account
does not permit to make a relative efficiency of the accretion lower
then $\sim 0.25$ from the standard value. It is expected that more accurate
treatement of the relaxation connected with the plasma turbulence will
even more increase the efficiency, making it close to unity (see also
Fabian and Rees, 1995).

Some observational data which were interpreted as an evidence for the
existence of the ADAF regime have disappered after additional accumulation
of data. The most interesting example of this sort is connected with the
claim of the proof of the existence of event horizon of the black holes
due to manifestation of the ADAF regime of accretion (Narayan et al., 1997).
Analysis of the more complete set of the observational data
(Chen et al., 1997) had shown disapperence of the statistical effect
clamed as an evidence for ADAF. This example shows how dangerous is to
base a proof of the theoretical model on the preliminary observational data.
It is even more dangerous, when the model is physically
not fully consistent. Then even a reliable set of the observational
data cannot serve as a proof of the model. The classical example from
astrophysics of this kind gives the theory of the origin of the elements
presented in the famous book of G.Gamov (1952), where the model of
the hot universe was developed. In addition to rich advantages of this model,
the author also wanted to explained
the origin of heavy elements in the primodial
explosion, neglecting the problems connected with an absence of the
stable elements with the number of barions equal to 5 and 8. G. Gamov
considered a good coinsidence of his calculations, where the mentioned
problem was neglected, and the observational curve, as a best proof of
his theory of the origin of the elements. The farther developements
have shown that his outstanding theory explains lot of things, exept
the origin of the heavy elements, which are produced due to stellar
evolution.

It looks like it is difficult to use ADAF for solution of
the problem of existence of
 underluminous AGN, where the observed flux of the energy is smaller,
 then the expected from the standard accretion disc models. Two possible
 ways may be suggested. One is based on a more accurate estimations
 of the accretion mass flow into the black hole, which could be
 overestimated. Another, more attractive possibility, is based
 on existence of another mechanisms of the energy losses in the form of
 accelerated particles, like in the radio-pulsars, where their losses
 exceed strongly a radiation losses. This is very probable to happen in
a presence of a large scale magnetic field which may be also responsible for
a fomation of the observed jets. To extend this line, we may  suggest, that
underlumilnous AGN loose main part of their enegry to the formation of jets.
The search of the correlation between existence of jets and lack of
the luminosity could be very informative.

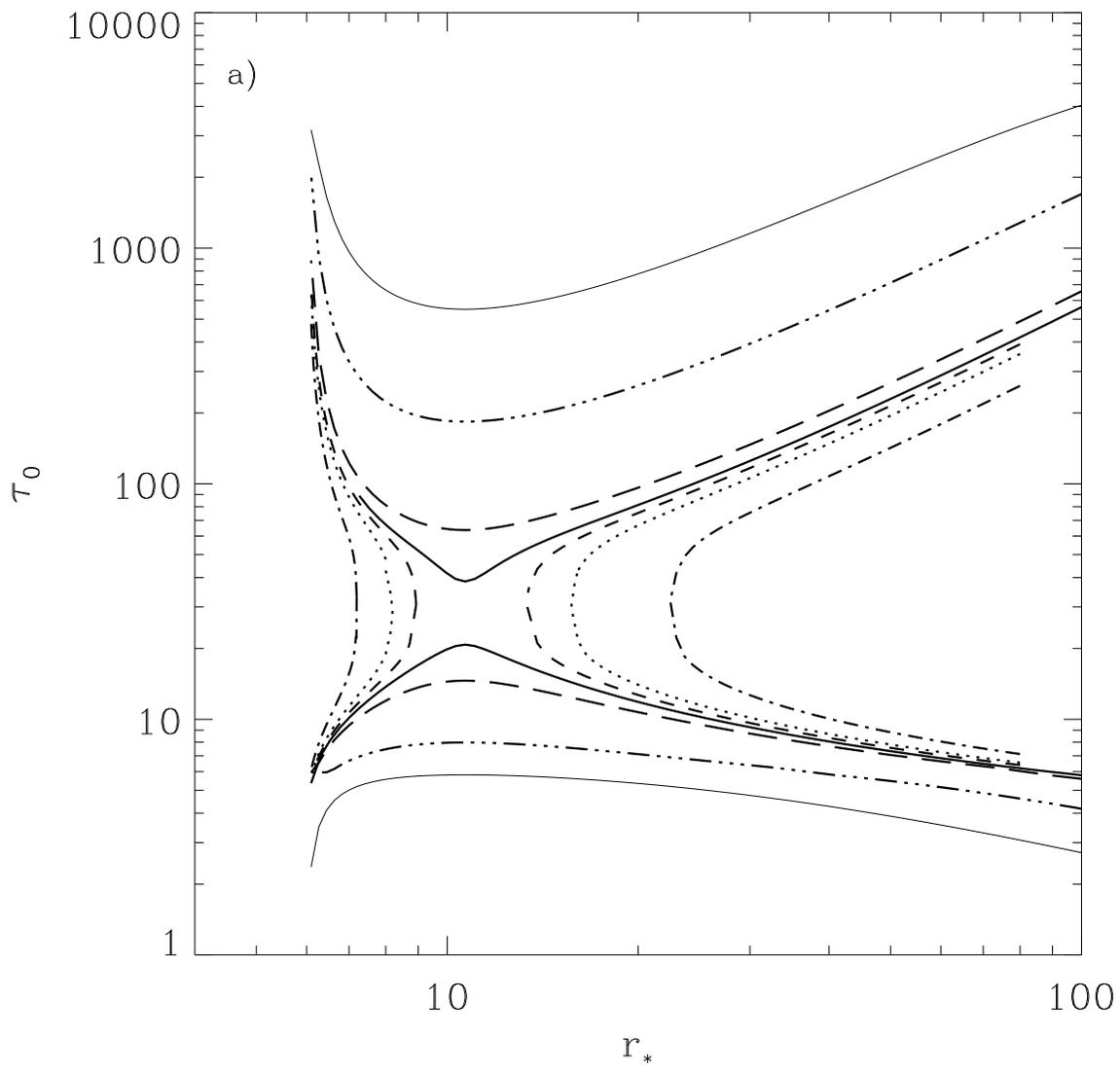
\end{document}